\documentclass[aps,prl,amsmath,amssymb,floatfix,twocolumn, amsmath, superscriptaddress, twocolumn]{revtex4}
\usepackage{multirow}
\usepackage{bbold}
\usepackage{subfigure}
\usepackage{color}
\usepackage{mathrsfs}
\usepackage{hyperref}
\usepackage[normalem]{ulem}
\usepackage{bm}

\usepackage{amsfonts, relsize, color}
\usepackage{graphics}
\usepackage{graphicx}
\usepackage{subfigure}
\usepackage{hyperref}
\usepackage{color}

\begin{document}
\title{Dissolution of topological Fermi arcs in a dirty Weyl semimetal}

\author{Robert-Jan Slager}
\affiliation{Max-Planck-Institut f$\ddot{\mbox{u}}$r Physik komplexer Systeme, N$\ddot{\mbox{o}}$thnitzer Str. 38, 01187 Dresden, Germany}

\author{Vladimir Juri\v ci\' c}
\affiliation{Nordita,   KTH Royal Institute of Technology and Stockholm University, Roslagstullsbacken 23,  10691 Stockholm,  Sweden}

\author{Bitan Roy}
\affiliation{Department of Physics and Astronomy, Rice University, Houston, Texas 77005, USA}
\affiliation{Condensed Matter Theory Center, University of Maryland, College Park, Maryland 20742, USA}
\affiliation{Max-Planck-Institut f$\ddot{\mbox{u}}$r Physik komplexer Systeme, N$\ddot{\mbox{o}}$thnitzer Str. 38, 01187 Dresden, Germany}

\begin{abstract}
Weyl semimetals (WSMs) have recently attracted a great deal of attention as they provide condensed matter realization of chiral anomaly, feature topologically protected Fermi arc surface states and sustain sharp chiral Weyl quasiparticles up to a critical disorder at which a continuous quantum phase transition (QPT) drives the system into a metallic phase. We here numerically demonstrate that with increasing strength of disorder the Fermi arc gradually looses its sharpness, and close to the WSM-metal QPT it completely dissolves into the metallic bath of the bulk. Predicted topological nature of the WSM-metal QPT and the resulting bulk-boundary correspondence across this transition can directly be observed in angle-resolved-photo-emmision-spectroscopy (ARPES) and Fourier transformed scanning-tunneling-microscopy (STM) measurements by following the continuous deformation of the Fermi arcs with increasing disorder in recently discovered Weyl materials.
\end{abstract}

\maketitle

\begin{figure}
\includegraphics[width=8.5cm,height=5.5cm]{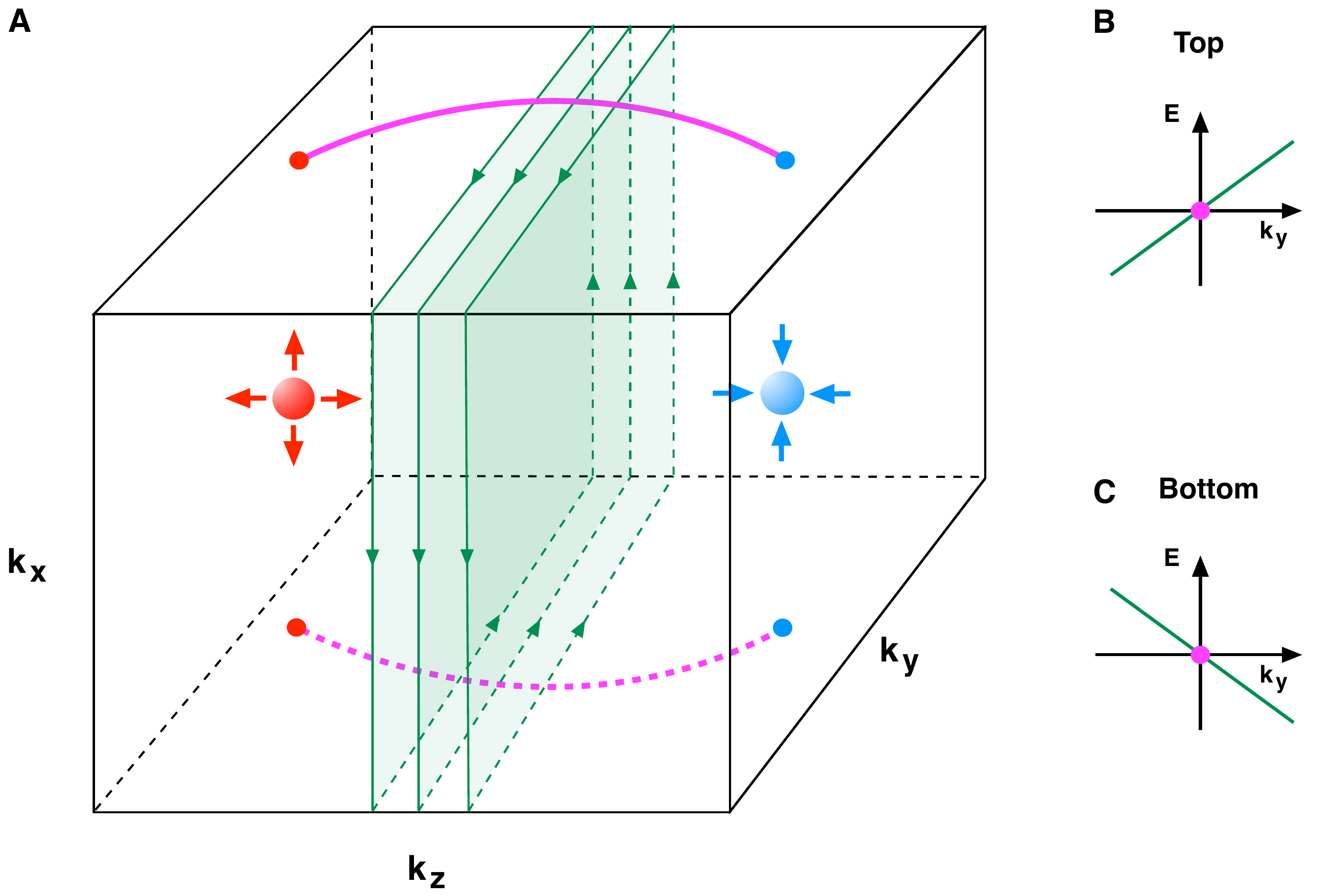}
\caption{ ({\bf A}) Topological origin of a Fermi arc [magenta lines on the top (solid) and bottom (dashed) surfaces] in a Weyl semimetal. Dispersion of one-dimensional chiral edge modes (dark green) ({\bf B}) on the top   and ({\bf C}) on the bottom surfaces, occupying the fraction of the surface Brillouin zone in between the projections of the left (red) and the right (blue) chiral Weyl points, which respectively act as a source (with outward arrows) and a sink (with inward arrows) of Abelian Berry curvature. The red and blue dots on the top and bottom surfaces are the projections of the bulk Weyl nodes. The Fermi arc is a locus of the zero-energy modes (magenta dots in {\bf B} and {\bf C}) of one-dimensional chiral edge states (shown in dark green with arrow heads in {\bf A}).
}~\label{FermiArc}
\end{figure}

\emph{Introduction.} With the rapid progress at the frontier of topological condensed matter physics, it has now become evident that the landscape of topological states of matter extends beyond gapped systems and Weyl semimetal (WSM) has emerged as the paradigmatic representative of a gapless topological phase~\cite{Schnyder,bernevig,hasan,armitage-2017}. It features chiral Weyl fermions as low energy excitations at pairs of points in the momentum space where the non-degenerate valence and conduction bands touch. These so called Weyl points, due to the lack of inversion and/or time-reversal symmetry, act as a source and sink of Abelian Berry curvature, and the monopole charge of the Weyl nodes defines the integer topological invariant of the system. Consequently, WSMs possess Fermi arcs as surface states that connect projections of the Weyl points onto the top and bottom surfaces, as illustrated in Fig.~\ref{FermiArc}.

Remarkably, a weakly disordered WSM describes a stable \emph{topological phase of matter}. On the other hand, at strong disorder, the WSM undergoes a quantum phase transition (QPT), beyond which Weyl fermions cease to exist as sharp quasiparticles and the system becomes a diffusive metal, as it has been recently established using both analytical~\cite{fradkin-main,shindou-murakami,chakravarty-first,balatsky-arovas,ominato,roy-dassarma-main,radzihovsky-main,altland,carpentier,JRS2016,shovkovy-main,lars,roy-dassarma-intdis,radzihovsky-2,juricic, GC-2,carpentier-2} and numerical~\cite{herbut-disorder-main,brouwer,pixley,chen-song-main,brouwer-2,hughes-main,roy-PRL,roy-bera,pixley-2,ohtsuki,fu-zhang} techniques. Concomitantly, the associated wide quantum critical regime supports a strongly coupled dirty non-Fermi liquid (NFL). However, the fate of the Fermi arc states in the vicinity of such QPT, a directly observable imprint of the associated NFL, has remained unexplored. We here address this problem of fundamental importance, which is intimately tied to the question of the topological nature of this transition, by numerically following the evolution of topologically protected Fermi arc states with increasing randomness in a WSM. We focus only on random charge impurities as they are the dominant source of elastic scattering in real materials.

\begin{figure*}[t]
\includegraphics[width=18cm,height=8cm]{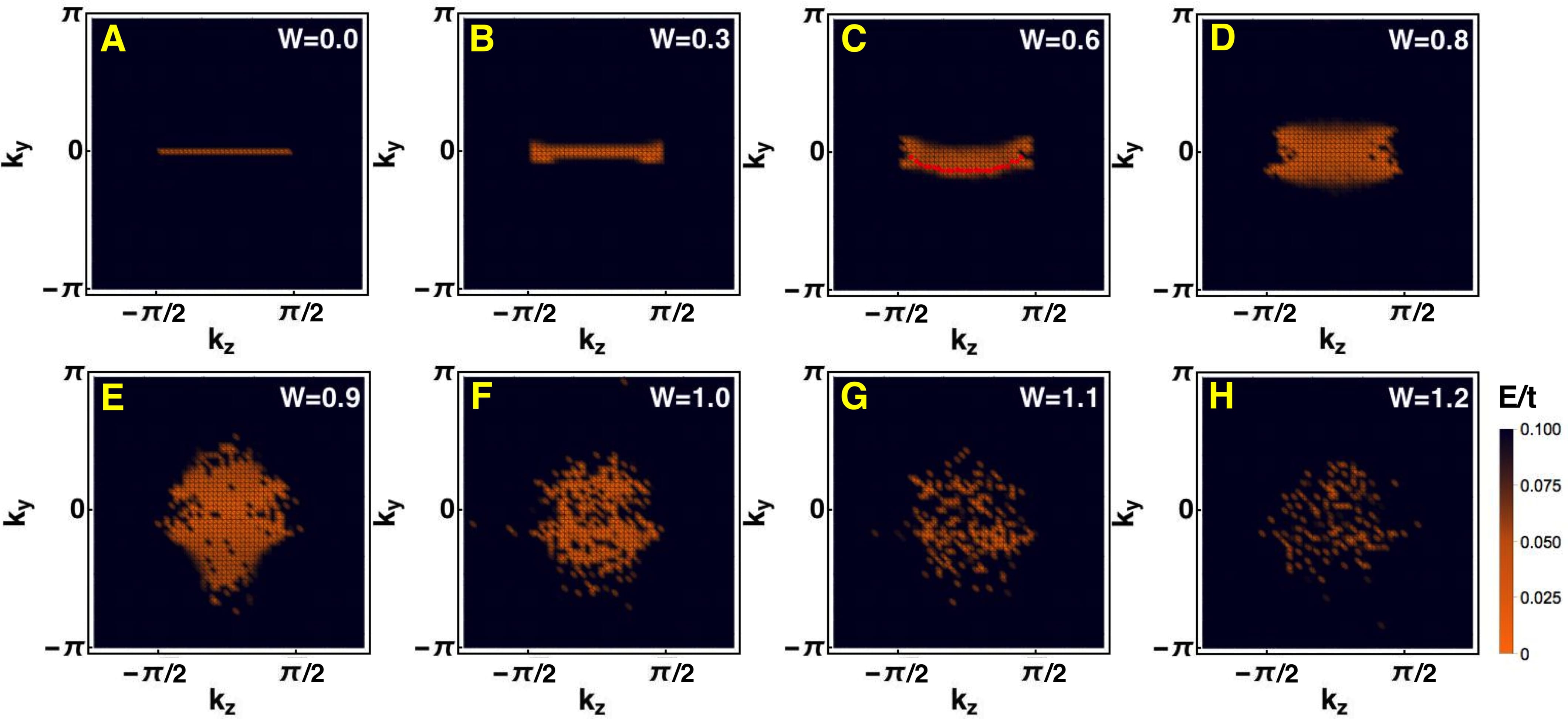}
\caption{ Evolution of one-dimensional chiral edge states connecting two Weyl points, located at $k_z=\pm \frac{\pi}{2 a}$  in the clean system, with increasing strength of disorder ($W$). The states shown are within the energy window $|E|/t \leq 0.1$ and  $a=1$ in the plots. As disorder increases,  more chiral edge states on the surface can be accommodated within a fixed energy window since the Fermi velocity of bulk Weyl fermions, and consequently the velocity of chiral edge modes, gradually decreases. For weak disorder the zero energy states can be joined to construct a sharp Fermi arc [red dotted line in {\bf C}]. But, at stronger disorder the number of states near zero energy and localized on the surface is rather sparse. In the proximity to the WSM-metal QPT that takes place for $W=W_c=1.0 \pm 0.1$ (Fig.~\ref{DOS_Bulk}) and inside the metallic phase, the topological Fermi arc gets dissolved into the metallic bulk and thus looses its support on the surface. The linear dimension in the $x$ direction (thickness) is $L=300$. Qualitatively similar results are found in systems with $L=100$ and $200$ [see Figs.~S2 and ~S3~\cite{supplementary}]
}~\label{energy-Fermiarc}
\end{figure*}

\begin{figure*}[t]
\includegraphics[scale=0.3]{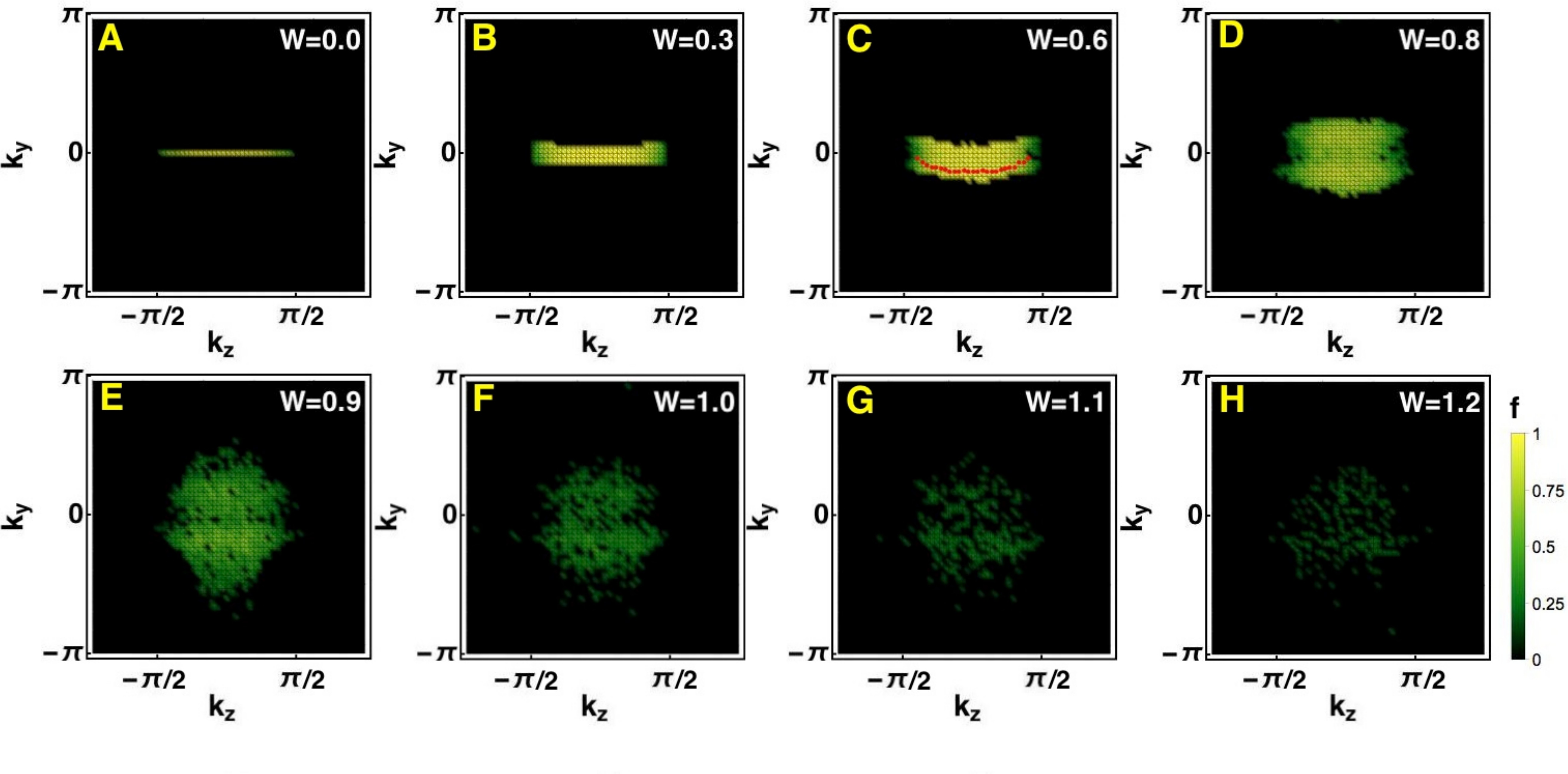}
\caption{ Fraction of the wave-function ($f$) localized on the surface for same set of low-energy chiral edge states shown in Fig.~\ref{energy-Fermiarc}. Even though at weak disorder the Fermi arc remains sharp [red dotted line in {\bf C}], it gradually dissolves into the emerging metallic bulk for stronger disorder. Thus, the disappearance of the Fermi arc promotes a bulk-boundary correspondence across the WSM-metal QPT. Qualitatively similar results are found in systems with $L=100$ and $200$, see Figs.~S4, S5~\cite{supplementary}. 
}~\label{localization-Fermiarc}
\end{figure*}

We demonstrate that topologically protected Fermi arc slowly dissolves into the emerging metallic bath accommodated by the bulk of a WSM as the strength of disorder gradually increases. At the brink of the onset of  metallicity the Fermi arc completely deliquesces, as the number of low energy states as well as the fraction of the wave function localized on the surface becomes extremely small near the WSM-metal QCP, shown in Figs.~\ref{energy-Fermiarc} and ~\ref{localization-Fermiarc}, which together constitute the central result of our work. 

Thus, a \emph{bulk-boundary correspondence} is unveiled across the WSM-metal QPT that also establishes the topological nature of this transition: The disappearance of the Fermi arcs is directly related to the vanishing of the bulk topological invariant, as the metallic phase, supporting quasiparticles with finite lifetime and mean-free path, is topologically trivial. The continuous dissolution of the Fermi arc can directly be observed through ARPES and Fourier transformed STM measurements in recently discovered WSM in weakly correlated materials (TaAs, TaP, NbP, etc.)~~\cite{taas-2,taas-3,nbas-1,tap-1,nbp-1,tas,felser-1,felser-2,yazdani-STM} and possibly in proposed WSMs in strongly correlated compounds, such as 227 pyrochlore iridates~\cite{vishwanath,goswami-roy}. Indeed, a recent tunneling microscopy experiment has shown robustness of the Fermi arc surface states in TaAs for weak disorder \cite{sessi-STM}, in agreement with our theoretical predictions.

\emph{Model.} To proceed with the numerical analysis, we subscribe to a simple realization of a WSM from the following tight-binding model on a cubic lattice
\begin{equation}~\label{latticemodel}
H ({\mathbf k})=t [ \sum_{j=x,y} \sigma_j \sin (k_x a) + \sigma_z [2-\sum_{j=x,y,z} \cos (k_j a) ] ]
\end{equation}
that features only two Weyl nodes at momenta $\pm {\mathbf K}_0$, where ${\mathbf K}_0=\left(0,0, \frac{\pi}{2 a}\right)$, with $a$ as the lattice spacing, ${\boldsymbol \sigma}$s as the standard Pauli matrices (see Fig.~S1 in Ref.~\cite{supplementary}), and $t=1$ and $\hbar=1$ set hereafter. In the vicinity of two Weyl nodes the low-energy excitations are described by left ($+$) and right ($-$) chiral Weyl fermions with the Hamiltonian
\begin{equation}
H_{\pm {\mathbf K}_0 + {\mathbf q}}= v \left( q_x \sigma_x +  q_y \sigma_y \pm q_z \sigma_z \right),
\end{equation}
where $v=t a$ bears the dimension of Fermi velocity. The low-energy dispersion around the Weyl nodes is given by $E_{\pm}(\mathbf q)=\pm v |{\mathbf q}|$, where $\pm$ corresponds to the conduction and valence band, respectively. Thus a WSM represents a fixed point in $d=3$ with $z=1$, where $z$ is the dynamic scaling exponent (DSE) that defines relative scaling between energy and momentum, according to $E\sim |{\mathbf q}|^z$. Consequently, the density of states (DOS) in a WSM scales as $\varrho(E) \sim |E|^2/v^3$, following the general scaling form $\varrho(E) \sim |E|^{-1+d/z}$.

\emph{Topology and the Fermi arc.} The topological origin of the Fermi arc surface state, illustrated in Fig.~\ref{FermiArc}A, can be demonstrated from the above tight-binding model as follows. If we set $k_z=0$, the resulting two-dimensional Hamiltonian corresponds to a lattice model for time-reversal symmetry breaking \emph{quantum anomalous Hall insulator}, which supports a topologically protected one-dimensional gapless edge state. Hence, a three-dimensional WSM can be envisioned as stacked layers of two-dimensional quantum anomalous Hall insulators in the $k_z$ direction, within the interval $-\frac{\pi}{2a} \leq k_z \leq \frac{\pi}{2a}$, for example, and each such layer accommodates a one-dimensional gapless chiral edge state. Thus, the resulting WSM also supports gapless edge states on the surface in the $k_z-k_y$ plane with energy dispersion $\pm v k_y$, respectively on the top and bottom surface, as shown in Figs.~\ref{FermiArc}B and ~\ref{FermiArc}C. The topologically protected Fermi arc is constituted by the \emph{locus} of the zero-energy states of these chiral modes between the two Weyl nodes, see Fig.~\ref{energy-Fermiarc}A. Notice that Fermi arc can also be found on the front and back faces in the $k_z-k_x$ plane. But, for concreteness, we here focus only on the top surface and therefore implement periodic boundary in the $y$ and $z$ directions. To expose the Fermi arc on the $k_z-k_y$ plane we impose open boundary in the $x$ direction, along which the linear dimension is denoted by $L$.

We note that the localization length (${\ell}$) of each such copy of chiral edge state is proportional to the bulk gap for a given value of $k_z$, and diverges ($\ell \to \infty$) as we approach the Weyl nodes, at $\pm {\mathbf K}_0$, from the center of the surface Brillouin zone ($k_z=0$). Consequently, the fraction of the wave-function localized on the surface (inversely proportional to $\ell$) decreases monotonically away from the center of the Fermi arc, as shown in Fig.~\ref{localization-Fermiarc}A.

\emph{Method.} We here introduce a new technique, referred as \emph{stacked-layer construction} [see Secs.~S1 and S2 of Ref.~\cite{supplementary} for details]: Disorder potential varies randomly and independently within the interval $[-W,W]$, but only along the $x$ direction, while it remains completely flat in each $yz$-plane. This method allows us to scan the evolution of the Fermi arc with disorder in  unprecedentedly large systems of $\sim 100^3$ lattice sites [Figs.~\ref{energy-Fermiarc}, \ref{localization-Fermiarc}, Ref.~\cite{supplementary}], and in turn to also probe the onset of a NFL phase.

\begin{figure}
\includegraphics[width=8.5cm, height=5.0cm]{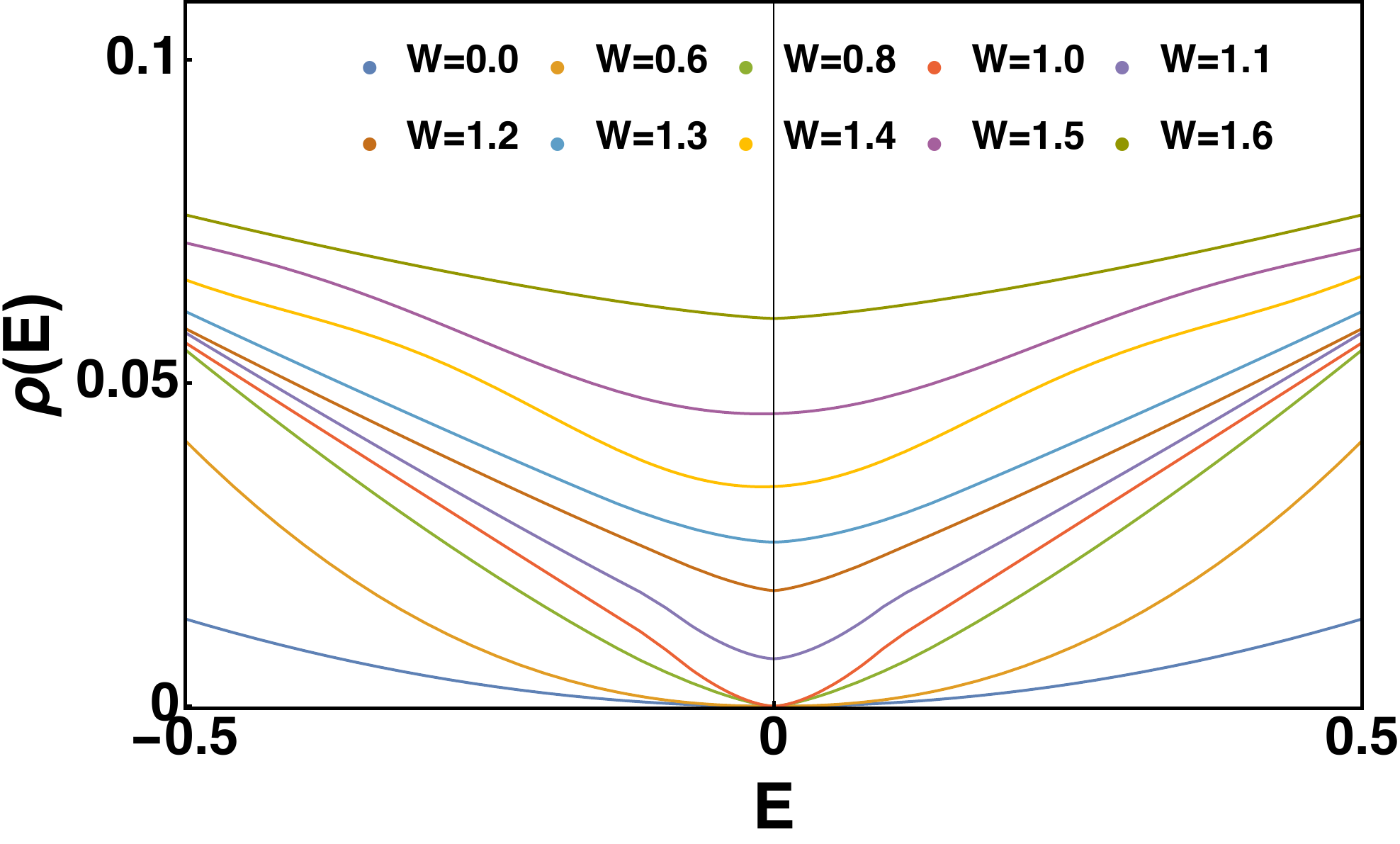}
\caption{ Average DOS $[\varrho(E)]$ in the bulk of a dirty WSM. The critical strength of disorder for WSM-metal QPT is $W_c = 1.0 \pm 0.1$; see Sec.~S2 for details~\cite{supplementary}. 
}~\label{DOS_Bulk}
\end{figure}

\emph{WSM-metal QPT.} To establish a bulk-boundary correspondence across the WSM-metal QPT, we first study the effects of random charge impurities in the bulk by computing the average DOS [$\varrho(E)$] in a dirty WSM with periodic boundary conditions in all three directions, as the average DOS at zero energy [$\varrho(0)$] allows us to numerically estimate the WSM-metal QCP~\cite{comment-rare-region}. The results displayed in Fig.~\ref{DOS_Bulk}, suggest that WSM remains a stable phase of matter up to a critical strength of disorder $W_c=1.0 \pm 0.1$, at which the system undergoes a continuous QPT and enters into a metallic phase, where $\varrho(0)$ becomes finite~\cite{supplementary}. Such QPT is characterized by DSE $z=1.49 \pm 0.05$~\cite{supplementary}, which matches quite well with the results in Refs.~\cite{herbut-disorder-main, pixley, pixley-2, ohtsuki, roy-bera} obtained by using kernel polynomial method~\cite{kpm}, and the one from scaling of conductance obtained via transfer matrix formalism~\cite{brouwer-2}. Consequently, the average DOS displays distinct power-law behavior in the WSM $\varrho(E) \sim |E|^2$ and at the WSM-metal QCP $\varrho(E) \sim |E|$, following its general scaling form. Furthermore, the Fermi velocity of Weyl fermion vanishes as the WSM-metal QCP is approached from the semimetallic side according to $v \sim |\delta|^{(z-1)\nu}$, where $\delta=(W-W_c)/W_c$. Such prediction is consistent with the feature that with increasing disorder a WSM becomes more metallic. Namely, DOS near zero energy increases without altering the power-law behavior for $W<W_c$, since $\varrho(E) \sim |E|^2/v^3$ before entering the metallic phase, where  $\varrho(0)$ becomes finite [Fig.~\ref{DOS_Bulk}] and scales as $\varrho(0) \sim \delta^{(d-z)\nu}$, with the correlation length exponent $\nu=1.02 \pm 0.05$~\cite{supplementary}, which agrees reasonably well with the ones obtained by using kernel polynomial method~\cite{herbut-disorder-main, pixley, pixley-2, ohtsuki, roy-bera}. Vanishing Fermi velocity indicates the lack of sharp chiral excitations at the WSM-metal QCP and the presence of a strongly coupled NFL inside the entire quantum critical regime at finite energies. We here extract these two exponents to benchmark a newly developed ``\emph{stacked-layer construction}" methodology. Notice the dispersion of chiral edge states $E=\pm v k_y$ is expected to become \emph{more flattened}, due to the decrease of the Fermi velocity in the bulk with increasing randomness. Next we numerically establish such qualitative change in the nature of the chiral edge states for weak disorder and breakdown of the notion of sharp Fermi arc states across the WSM-metal QPT.

\emph{Bulk-boundary correspondence.} Fig.~\ref{energy-Fermiarc} displays the evolution of one-dimensional chiral edge states within a fixed energy window on the top surface for weak ($W<W_c$) and strong ($W>W_c$) disorder. As the Fermi velocity of bulk Weyl quasiparticles decreases with increasing disorder, a larger number of chiral edge states can be accommodated within the same energy window at weak disorder, as depicted in Fig.~\ref{energy-Fermiarc}. However, as the strength of disorder approaches the critical one for WSM-metal QPT in the bulk, the number of low-energy states localized on one surface decreases continuously. Finally, in the close proximity to the QCP as well as inside the metallic phase only very few low-energy states are localized on the surface and the notion of a sharp Fermi arc becomes moot. Thus across the WSM-metal QPT the Fermi arc completely dissolves into the bulk metallic background.

To further anchor the topological nature of the WSM-metal QPT, in Fig.~\ref{localization-Fermiarc} we display the fraction of the wave-function for the same low-energy surface states as shown in Fig.~\ref{energy-Fermiarc}. With increasing disorder the fraction of the wave function associated with the low-energy surface states decreases monotonically. Close to the WSM-metal QPT, as well as inside the metallic phase, only a tiny fraction of the wave-function is localized on the surface, consistent with our previous observation that at strong disorder topologically protected Fermi arc looses its support on the surface. Thus, our numerical analysis unambiguously establishes a direct bulk-boundary correspondence across the WSM-metal QCP and corroborates in favor of the topological nature of this transition.

Typically we find that Fermi arc dissolves for $W=W_\ast<W_c$, with $W_\ast \approx 0.8$ [Figs.~\ref{localization-Fermiarc} and ~\ref{DOS_Bulk}], irrespective (approximately) of the system size~\cite{supplementary}. In any finite system arc states from opposite surfaces enjoy a finite overlap, which is inversely proportional to the localization length ($\ell$) of the arc state for a given value of $k_z$. Thus, such an overlap is nominal at the center of the arc, but increases monotonically as we approach the Weyl node, where the arc states from the top and bottom surfaces get connected via the bulk Weyl points. However, disorder introduces a new length scale in the system, the \emph{correlation length} ($\xi$), which diverges as $\xi \sim |\delta|^{-\nu}$ in the vicinity of the WSM-metal QCP. When $\xi \sim L$ the entire Fermi arcs from opposite faces overlap significantly and they loose support on the surface. Quite naturally, in any finite system this occurs for $W<W_c$, and the Fermi arc dissolves into the metallic bath of a Weyl system for a subcritical strength of disorder. In addition, the end points of the Fermi arc where overlap between the two surfaces is much larger disappear for even slightly weaker disorder ($W<W_\ast$). Hence, the dissolution of a sharp Fermi arc at strong disorder is a precursor of the WSM-metal QPT in the bulk, which through bulk-boundary correspondence unveils the topological nature of this transition. Furthermore, when $\xi \sim L$ the system enters the critical regime associated with the WSM-metal QPT that supports a strongly coupled NFL lacking sharp quasi-particle excitations. Thus, dissolution of the Fermi arc, besides the bulk-boundary correspondence, stands as a fingerprint of an underlying NFL phase of matter. 
At finite temperature the Fermi arc retains its sharpness if $\xi \ll \lambda_{th}$, where $\lambda_{th} \sim \frac{\hbar v}{k_B T}$ is the thermal de-Broglie wavelength. In a finite system the Fermi arc dissolves when $\xi \sim \mbox{min} (L, \lambda_{th})$. Our results regarding the dissolution of ``\emph{shortened}" Fermi arcs (now restricted within $-\frac{\pi}{2a}+ |\frac{\mu}{v}| \leq k_z \leq \frac{\pi}{2a}- |\frac{\mu}{v}|$) remain operative when chemical potential ($\mu$) is placed slightly away from Weyl nodes.

\emph{Discussion.} From the gradual disappearance of a sharp Fermi arc on the surface of an increasingly disordered WSM, we reveal the bulk-boundary correspondence associated with the WSM-metal QPT as well its topological nature that can also be observed in various Weyl materials through ARPES and Fourier transformed STM measurements. Such an outcome is also germane for Weyl superconductors and superfluids, e.g. in the $A$ phase of $^3$He~\cite{volovik} and doped half-Heusler compounds~\cite{roy-nevidomskyy}, as well as for topological Dirac semimetals, realized in Cd$_3$As$_2$ and Na$_3$Bi~\cite{cdas, nabi, nabi-arc}, since they feature linearly dispersing quasiparticles in the bulk and arc states on the surface. Our results open up a route to investigate the fate of the Fermi arcs in other members of the Weyl family, such as double and triple WSMs, in the presence of randomness. As disorder is respectively marginally relevant and relevant in these two systems~\cite{roy-bera}, we expect the Fermi arcs to disappear for infinitesimal strength of disorder. Finally, our findings should motivate future investigations on the role of other types of disorder~\cite{juricic}, various non-perturbative effects~\cite{nandkishore-rare, pixley-huse, pixley-nandkishore, gurarie, ostrovsky, fu-zhang}, such as puddles, rare regions, and  Lifshitz tails \cite{RMP-transport-graphene}, on the topological nature of WSM-metal QPT and Fermi arcs, as well as stability of Fermi arcs against disorder deposited only on the surface~\cite{surface-comment}.

\emph{Acknowledgments.} B. R. was supported by Welch Foundation Grant No.~C-1809, NSF CAREER grant no.~DMR-1552327 of Matthew S. Foster. We are thankful to Matthew Foster, Pallab Goswami, Subir Sachdev, Jay D. Sau for useful discussions. R-J. S. and B. R. are thankful to Nordita, Center for Quantum Materials for hospitality.

{}


\begin{thebibliography}{}

\bibitem{Schnyder} C.-K Chiu, J. C. Y.  Teo,  A. P. Schnyder, and S. Ryu,  { Rev. Mod. Phys.} {\bf 88}, 035005 (2016).

\bibitem{bernevig} B. Bradlyn, J. Cano, Z. Wang, M. G. Vergniory, C. Felser, R. J. Cava, and B. A. Bernevig, 
{ Science} {\bf 353}, 558 (2016).

\bibitem{hasan} M.Z. Hasan, S.-Y. Xu, I. Belopolski, and S.-M. Huang,  { Ann. Rev. Cond. Matt. Phys.} {\bf 8}, 289 (2017).    

\bibitem{armitage-2017} N.P. Armitage, E.J.  Mele, and A. Vishwanath, arXiv:1705.01111 (2017).


\bibitem{fradkin-main} E. Fradkin,  { Phys. Rev. B} {\bf 33}, 3263 (1986).

 \bibitem{shindou-murakami} R. Shindou, and S. Murakami,{ Phys. Rev. B} {\bf 79}, 045321 (2009).

\bibitem{chakravarty-first} P. Goswami, and S. Chakravarty, { Phys. Rev. Lett.} {\bf 107}, 196803 (2011).

\bibitem{balatsky-arovas}  Z. Huang,T. Das,  A.V. Balatsky, and  D. P. Arovas,  { Phys. Rev. B} {\bf 87}, 155123 (2013).

\bibitem{roy-dassarma-main} B. Roy, and S. Das Sarma, { Phys. Rev. B} {\bf 90}, 241112(R) (2014).

\bibitem{ominato} Y. Ominato and M. Koshino, Phys. Rev. B {\bf 89}, 054202 (2014).

\bibitem{radzihovsky-main} S.V. Syzranov, L. Radzihovsky, and V. Gurarie, { Phys. Rev. Lett.} {\bf 114}, 166601 (2015).

\bibitem{altland} A. Altland and D. Bagrets, { Phys. Rev. Lett.} {\bf 114}, 257201 (2015).

\bibitem{carpentier}  T. Louvet, D. Carpentier, and A.A. Fedorenko,  { Phys. Rev. B} {\bf 94}, 220201 (2016).

\bibitem{JRS2016} B. Roy,  V. Juri\v ci\' c, and S. Das Sarma, { Sci. Rep.} {\bf 6}, 32446 (2016).

\bibitem{shovkovy-main} E.V. Gorbar, V.A. Miransky, I. A. Shovkovy, and P. O. Sukhachov, { Phys. Rev. B} {\bf 93}, 235127 (2016).

\bibitem{lars} A. K. Mitchell,  and L. Fritz,  { Phys. Rev. B} {\bf 93}, 035137 (2016).

\bibitem{roy-dassarma-intdis} B. Roy and S. Das Sarma, Phys. Rev. B {\bf 94}, 115137 (2016).

\bibitem{radzihovsky-2}  S. V. Syzranov, V. Gurarie, and L. Radzihovsky,  Ann. Phys. {\bf 373}, 694 (2016)

\bibitem{juricic} B. Roy, R.-J. Slager, and V. Juri\v ci\' c,  arXiv:1610.08973 (2016).

\bibitem{GC-2} P. Goswami and S. Chakravarty, Phys. Rev. B {\bf 95}, 075131 (2017).

\bibitem{carpentier-2} T. Louvet, D. Carpentier, and A. A. Fedorenko, Phys. Rev. B {\bf 95}, 014204 (2017).


\bibitem{herbut-disorder-main} K. Kobayashi, T. Ohtsuki, K.-I. Imura,  and I. F. Herbut,  { Phys. Rev. Lett.} {\bf 112}, 016402 (2014).

\bibitem{brouwer} B. Sbierski, G. Pohl, E. J.  Bergholtz, and P. W. Brouwer,  { Phys. Rev. Lett.} {\bf 113}, 026602 (2014).

\bibitem{pixley} J. H.  Pixley, P.  Goswami, and S. Das Sarma, { Phys. Rev. Lett.} {\bf 115}, 076601 (2015).

\bibitem{chen-song-main} C.-Z Chen, J. Song, H. Jiang, Q.-f. Sun, Z. Wang, and X. C. Xie  { Phys. Rev. Lett.} {\bf 115}, 246603 (2015).

\bibitem{brouwer-2} B. Sbierski, E. J. Bergholtz, and P. W. Brouwer, { Phys. Rev. B} {\bf 92}, 115145 (2015).

\bibitem{hughes-main} H. Shapourian, and T. L. Hughes, { Phys. Rev. B} {\bf 93}, 075108 (2016).

\bibitem{roy-PRL} B. Roy, Y. Alavirad, and J. D. Sau, Phys. Rev. Lett. {\bf 118}, 227002 (2017).

\bibitem{roy-bera} S. Bera, J. D. Sau, and B. Roy,  { Phys. Rev. B} {\bf 93}, 201302 (2016).

\bibitem{pixley-2} J.H. Pixley, P.  Goswami, and S. Das Sarma, { Phys. Rev. B} {\bf 93}, 085103 (2016).

\bibitem{ohtsuki} S. Liu, T.  Ohtsuki, and R. Shindou, { Phys. Rev. Lett.} {\bf 116}, 066401 (2016).

\bibitem{fu-zhang} B. Fu, W. Zhu, Q. Shi, Q. Li, J. Yang, and Z. Zhang,  Phys. Rev. Lett. {\bf 118}, 146401 (2017).


\bibitem{taas-2}  S.-Y. Xu, I. Belopolski, N. Alidoust, M. Neupane, G. Bian, C. Zhang, R. Sankar, G. Chang, Z. Yuan, C.-C. Lee, 
S.-M. Huang, H. Zheng, J. Ma, D. S. Sanchez, B.K. Wang, A. Bansil, F. Chou, P. P. Shibayev, H. Lin, S. Jia, and M. Z. Hasan, 
{ Science} {\bf 349}, 613 (2015).

\bibitem{taas-3}  B. Q. Lv, H. M. Weng, B. B. Fu, X. P. Wang, H. Miao, J. Ma, P. Richard, X. C. Huang, L. X. Zhao, G. F. Chen, 
Z. Fang, X. Dai, T. Qian, and H. Ding, { Phys. Rev. X} {\bf 5}, 031013 (2015).

\bibitem{nbas-1}  S.-Y. Xu, N. Alidoust,	I. Belopolski,	Z. Yuan,	G. Bian,	T.-R. Chang, H. Zheng,	V. N. Strocov,	D. S. Sanchez,	G. Chang,	C. Zhang, D. Mou,	Y. Wu,	L. Huang,	C.-C. Lee,	S.-M. Huang,	B. Wang, A. Bansil,	H.-T. Jeng,	T. Neupert,	A. Kaminski,	H. Lin,	
S. Jia, and M. Zahid Hasan , { Nat. Phys.} {\bf 11}, 748 (2015).

\bibitem{felser-1}  L. X. Yang,	Z. K. Liu,	Y. Sun,	H. Peng,	H. F. Yang,	T. Zhang,	B. Zhou,	Y. Zhang,	Y. F. Guo,	M. Rahn,	D. Prabhakaran,	Z. Hussain,	S.-K. Mo,	C. Felser,	B. Yan, and Y. L. Chen, { Nat. Phys.} {\bf 11}, 728 (2015).

\bibitem{nbp-1}  C. Shekhar, A. K. Nayak,	Y. Sun,	M. Schmidt,	M. Nicklas,	I. Leermakers,	U. Zeitler,	Y. Skourski,	J. Wosnitza,	Z. Liu,	Y. Chen,	W. Schnelle,	H. Borrmann,	Y. Grin,	C. Felser, and Binghai Yan, { Nat. Phys.} {\bf 11}, 645 (2015).

\bibitem{tap-1} N. Xu, H. M. Weng, B. Q. Lv, C. E. Matt, J. Park, F. Bisti, V. N. Strocov, D. Gawryluk, E. Pomjakushina, K. Conder, N. C. Plumb, M. Radovic, G. Autes, O. V. Yazyev, Z. Fang, X. Dai, T. Qian, J. Mesot, H. Ding, and M. Shi, { Nat. Commun.} {\bf 7}, 11006 (2016).

\bibitem{tas} G. Chang, G. Chang, S.-Y. Xu3, D. S. Sanchez, S.-M. Huang, C.-C. Lee, T.-R. Chang, G. Bian, H. Zheng, I. Belopolski, 
N. Alidoust, H.-T. Jeng, A. Bansil, H. Lin, and M. Z. Hasan, { Sci. Adv.} {\bf 2}, e1600295 (2016).

\bibitem{yazdani-STM} H. Inoue, A. Gyenis, Z. Wang, J. Li, S. W. Oh, S. Jiang, N. Ni, B. A. Bernevig, and A. Yazdani, { Science} {\bf 351}, 1184 (2016).

\bibitem{felser-2} R. Batabyal,  N. Morali, N. Avraham, Y. Sun, M. Schmidt, C. Felser, A. Stern, B. Yan, and H. Beidenkopf, 
{ Sci. Adv.} {\bf 2}, e1600709 (2016).


\bibitem{vishwanath} X. Wan, A.M. Turner,  A.Vishwanath, and S. Y. Savrasov, Phys. Rev. B {\bf 83}, 205101 (2011).

\bibitem{goswami-roy} P. Goswami, B.  Roy, and S. Das Sarma, Phys. Rev. B {\bf 95}, 085120 (2017).


\bibitem{sessi-STM} P. Sessi, Y. Sun, T. Bathon, F. Glott, Z. Li, H. Chen, L. Guo, X. Chen, M. Schmidt, C. Felser, B. Yan, 
and M. Bode, { Phys. Rev. B} {\bf 95}, 035114 (2017).

\bibitem{supplementary} Methodology and additional numerical results are available in online Supplementary Materials.


\bibitem{comment-rare-region} Note that $\varrho(0)$ serves as an order-parameter for WSM-metal QPT, up to non-perturbative effects such as rare-regions, puddles Lifshitz-tail etc., leading to finite, but \emph{extremely small} $\varrho(0)$ (beyond our numerical resolution) for weak enough randomness~\cite{nandkishore-rare, pixley-huse, pixley-nandkishore, gurarie, ostrovsky}. However, onset of genuine \emph{metallicity} from such non-perturbative effects can only be confirmed from non-zero typical DOS at $E=0$ and finite dc conductivity as $T \to 0$ [see \emph{50 Years of Anderson Localization}, ed. by E. Abrahams (World Scientific Publishing Company, 2010)], yet to be demonstrated. 


\bibitem{kpm} A. Wei\ss{}e, G.  Wellein, A. Alvermann, and H. Fehske,  { Rev. Mod. Phys.} {\bf 78}, 275 (2006).

\bibitem{volovik} G. E. Volovik, {\it The Universe in a Helium droplet.} Clarendon Press, Oxford (2003).

\bibitem{roy-nevidomskyy} B. Roy, S. A. A. Ghorashi, M. S. Foster and A. H. Nevidomskyy, arXiv:1708.07825

\bibitem{cdas}  S. Borisenko, Q. Gibson, D. Evtushinsky, V. Zabolotnyy, B. Buchner, and R. J. Cava, {\ Phys. Rev. Lett.} {\bf 113}, 027603 (2014).

\bibitem{nabi} Z.K. Liu, B. Zhou, Y. Zhang, Z. J. Wang, H. M. Weng, D. Prabhakaran, S.-K. Mo,
Z. X. Shen, Z. Fang, X. Dai, Z. Hussain, and Y. L. Chen, {\ Science} {\bf 343}, 864 (2014).

\bibitem{nabi-arc}  S.-Y. Xu, C. Liu, S. K. Kushwaha, R. Sankar, J. W. Krizan, I. Belopolski, M. Neupane, G. Bian, N. Alidoust, 
T.-R. Chang, H.-T. Jeng, C.-Y. Huang, W.-F. Tsai, H. Lin, P. P. Shibayev, F.-C. Chou, R. J. Cava, and M. Z. Hasan, { Science} {\bf 347}, 294 (2015).






\bibitem{nandkishore-rare} R. Nandkishore, D. A. Huse, and S.L. Sondhi, { Phys. Rev. B} {\bf 89}, 245110 (2014).

\bibitem{pixley-huse} J. H. Pixley,  D. A. Huse,  and S. Das Sarma,{ Phys. Rev. X} {\bf 6}, 021042 (2016).

\bibitem{pixley-nandkishore} J. H. Pixley, Y-Z. Chou, P. Goswami, D. A. Huse, R. Nandkishore, L. Radzihovsky, and S. Das Sarma,
Phys. Rev. B {\bf 95}, 235101 (2017).

\bibitem{gurarie} V. Gurarie, Phys. Rev. B {\bf 96}, 014205 (2017).

\bibitem{ostrovsky} T. Holder, C-W. Huang, P. Ostrovsky, arXiv:1704.05481

\bibitem{RMP-transport-graphene} S. Das Sarma, S. Adam, E. H. Hwang, and E. Rossi, { Rev. Mod. Phys.} {\bf 83}, 407 (2011).



\bibitem{surface-comment} Since in our setup disorder is randomly distributed in the bulk and on the surface of the system, we can safely conclude that at least for sufficiently weak disorder, deposited only on the surface, the Fermi arc remains stable, in qualitattive agreement with Ref.~\cite{sessi-STM}.   



\end{thebibliography}
\end{document}